\documentclass[aps,twocolumn,superscriptaddress]{revtex4-1}

\usepackage{amsmath}
\usepackage{amssymb}
\usepackage{amsthm}
\usepackage{graphicx}
\usepackage{hyperref}
\usepackage{xcolor}

\newcommand{\ket}[1]{| #1 \rangle}

\begin{document}

\title{Non-Gaussian entanglement criteria for atomic homodyne detection}

\author{Jaehak Lee}
\email{jaehak.lee.201@gmail.com}
\affiliation{School of Computational Sciences, Korea Institute for Advanced Study, Seoul 02455, Korea}
\author{Jiyong Park}
\affiliation{School of Basic Sciences, Hanbat National University, Daejeon 34158, Korea}
\author{Jaewan Kim}
\affiliation{School of Computational Sciences, Korea Institute for Advanced Study, Seoul 02455, Korea}
\author{M. S. Kim}
\affiliation{School of Computational Sciences, Korea Institute for Advanced Study, Seoul 02455, Korea}
\affiliation{QOLS, Blackett Laboratory, Imperial College London, London SW7 2AZ, United Kingdom}
\author{Hyunchul Nha}
\email{hyunchul.nha@qatar.tamu.edu}
\affiliation{Department of Physics, Texas A \& M University at Qatar, P.O. Box 23874, Doha, Qatar}

\begin{abstract}
Homodyne measurement is a crucial tool widely used to address continuous variables for bosonic quantum systems. While an ideal homodyne detection provides a powerful analysis, e.g. to effectively measure quadrature amplitudes of light in quantum optics, it relies on the use of a strong reference field, the so-called local oscillator typically in a coherent state. Such a strong coherent local oscillator may not be readily available particularly for a massive quantum system like Bose-Einstein condensate (BEC), posing a substantial challenge in dealing with continuous variables appropriately. It is necessary to establish a practical framework that includes the effects of non-ideal local oscillators for a rigorous assessment of various quantum tests and applications. We here develop entanglement criteria beyond Gaussian regime applicable for this realistic homodyne measurement that do not require assumptions on the state of local oscillators. We discuss the working conditions of homodyne detection to effectively detect non-Gaussian quantum entanglement under various states of local oscillators.
\end{abstract}

\maketitle

\section{\label{sec:introduction}Introduction}

In recent decades, continuous-variable (CV) quantum information \cite{Weedbrook2012,Serafini2017} has been developed to provide wide applications, e.g. quantum cryptography \cite{Pirandola2020}, quantum metrology \cite{Pirandola2018}, quantum computation \cite{Menicucci2006}, {\it etc}. In CV quantum information processing, continuous variables like position and momentum for a massive particle or the quadrature amplitude of light are employed to encode information and homodyne measurement provides a critical tool to measure these observables \cite{Serafini2017,Lvovsky2009}. It has been adopted as a powerful tool in various areas of CV quantum information processing including quantum teleportation \cite{Yonezawa2004,Pirandola2015}, quantum communication \cite{Takeoka2014,Lee2015,Lee2016}, quantum key distribution \cite{Grosshans2003}, and nonclassicality detection \cite{Park2017,Park2019,Park2021PRR}.

In an ideal homodyne measurement, a local oscillator (LO) of classical nature is required usually in a coherent state with a very large amplitude. In optical systems, an intense coherent state is readily produced using a strong laser field. On the other hand, if CV quantum information processing is to be implemented in other experimental platforms, such a LO may not be available in a desired form, e.g., for massive systems like Bose-Einstein condensates (BEC). Homodyne measurements in atomic systems have been realized experimentally to detect CV quantum correlations of massive particles to some extent \cite{Gross2011,Peise2015}. Recently, it was proposed that one can detect quantum gravity by observing non-Gaussianity of BEC using homodyne measurement \cite{Howl2021}. However, there exist some caveats when implementing homodyne measurements in atomic systems. First, the intensity strength of LOs is limited by the number of atoms available in a certain electronic state used as a reference state for LO, unlike the large number of photons in optical systems. Second, atomic ensembles do not remain as stable coherent states due to atomic interactions. There have been only few studies examining the effect of these imperfect LOs in addressing continuous variable for a massive system \cite{Ferris2008}.

In this paper, we develop entanglement criteria that can be tested via homodyne detection for a massive system by incorporating the properties of LO states appropriately. The existing entanglement criteria via quadrature observables \cite{Duan2000,Simon2000,Walborn2009,Nha2012} cannot be directly used when the strong oscillator limit is not satisfied. In this case, the actually measured observables are dependent on the properties of LOs as well as those of the signal fields under test \cite{Ferris2008}, thus requiring need for a careful approach to consider the effect of LOs. We here examine Hillery-Zubairy (HZ) criteria \cite{Hillery2006PRL,Hillery2006PRA} in the case that the assumption on classical LOs is not applicable. Remarkably, we show that the only information required about LOs is the mean number of bosons, which can be readily measured in a typical absorption-based imaging. We apply our criteria to specific examples of different LO states and examine how effectively they detect entanglement with different states of LOs.

This paper is organized as follows. In Sec. \ref{sec:homodyne}, we briefly introduce homodyne measurements and discuss how to deal with imperfect LOs. In Sec. \ref{sec:firstcriteria}, we show how to test first order HZ criteria using homodyne measurement and generalize it to the case of practical homodyne measurements. In Sec. \ref{sec:highercriteria}, we show how to also derive higher-order entanglement criteria incorporating realistic homodyne detections and discuss their resource requirements. In Sec. V, we analyze our criterion in a practical BEC system to illustrate the usefulness of our approach and conclude with remarks in Sec. \ref{sec:discussion}.

\section{\label{sec:homodyne}Homodyne measurement}

\begin{figure}[t]
	\centering \includegraphics[clip=true, width=0.6\columnwidth]{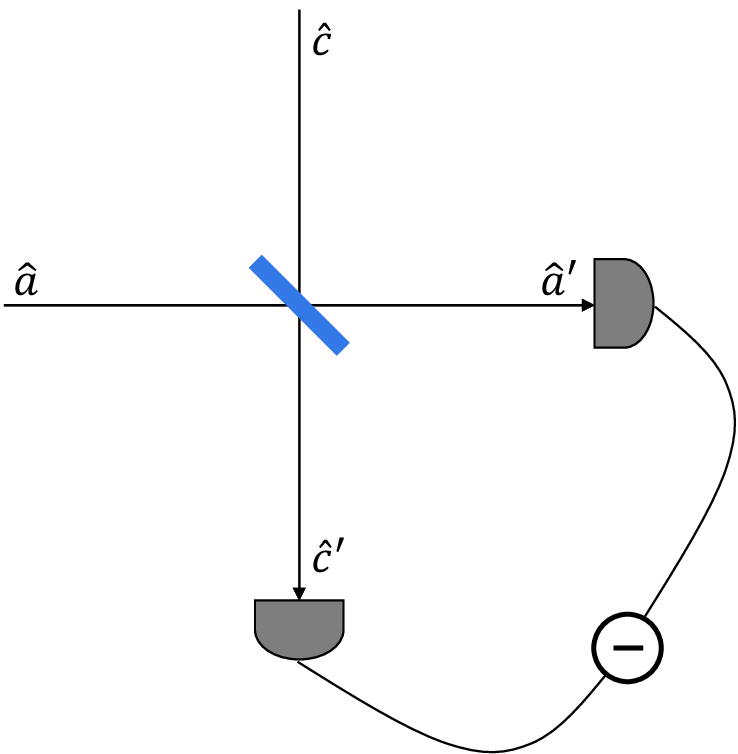}
    	\caption{\label{fig:homodyne} Scheme implementing a homodyne measurement. A signal mode $ \hat{a} $ and a LO mode $ \hat{c} $ are mixed at a 50/50 beam splitter. The difference in the intensity between two output modes provides information on the quadrature observables defined in main text.}
    \end{figure}

Bosonic quantum systems may be described by the annihilation and the creation operators $ \hat{a} $ and $ \hat{a}^\dagger $. They satisfy the commutation relation $ [\hat{a},\hat{a}^\dagger] = 1 $ and the bosonic number operator is given by $ \hat{n}_a = \hat{a}^\dagger\hat{a} $. In CV quantum information, they are alternatively described by quadrature operators such as position $ \hat{X}_a = \frac{1}{2}(\hat{a}^\dagger+\hat{a}) $ and momentum $ \hat{P}_a = \frac{i}{2}(\hat{a}^\dagger-\hat{a}) $. Quadrature operators can be measured by homodyne measurements that are generally implemented as shown in Fig. \ref{fig:homodyne}.
The signal mode $ \hat{a} $ is mixed with a LO $ \hat{c} $ at a 50/50 beamsplitter, then the operators associated with output modes may be written as $ \hat{a}' = \frac{1}{\sqrt{2}}(\hat{a}+\hat{c}) $ and $ \hat{c}' = \frac{1}{\sqrt{2}}(\hat{a}-\hat{c}) $. We measure the intensity of each output mode and consider the difference that corresponds to
    \begin{eqnarray}
	\hat{a}'^\dagger\hat{a}' - \hat{c}'^\dagger\hat{c}' & = & \frac{1}{2}(\hat{a}^\dagger+\hat{c}^\dagger)(\hat{a}+\hat{c}) - \frac{1}{2}(\hat{a}^\dagger-\hat{c}^\dagger)(\hat{a}-\hat{c}) \nonumber \\
	& = & \hat{a}^\dagger\hat{c} + \hat{a}\hat{c}^\dagger .
    \end{eqnarray}
In the strong oscillator limit, where the LO is in coherent state $ \ket{\alpha} $ with $\alpha=|\alpha|e^{\phi}$ and $ |\alpha| \gg 1 $, the difference signal is approximated as
    \begin{equation}
	\hat{a}^\dagger\hat{c} + \hat{a}\hat{c}^\dagger \approx |\alpha|\left(\hat{a}^\dagger e^{-\phi} + \hat{a}e^{\phi}\right) \equiv 2|\alpha|\hat{Q}_{a,\phi}.
    \end{equation}
Here a general quadrature $\hat{Q}_{a,\phi}$ is defined as $\hat{Q}_{a,\phi}\equiv\frac{\hat{a}^\dagger e^{-\phi} + \hat{a}e^{\phi}}{2}$. Therefore, depending on the phase $\phi$ of the LO, we can measure all different quadrature amplitudes. For instance, with $\phi=0$, the output signal corresponds to the position operator $ \hat{X}_a $. On the other hand, with $\phi=\frac{\pi}{2}$, we obtain the momentum operator $ \hat{P}_a $.
	
However, if the LO is not in the assumed coherent state of strong intensity, the above analysis does not hold and we must consider the measurement statistics more carefully. That is, the obtained signal of intensity difference corresponds to
    \begin{equation} \label{eq:Xm}
	\hat{X}_a^{(m)} \equiv \frac{\hat{a}^\dagger\hat{c} + \hat{a}\hat{c}^\dagger}{2\sqrt{\langle \hat{c}^\dagger\hat{c} \rangle}},
    \end{equation}
incorporating the normalization with respect to the finite intensity $\langle \hat{c}^\dagger\hat{c}\rangle$ of the LO. 
From now on, we use the superscript $ (m) $ to represent the actually ``measured'' observable distinguished from the ideal quadrature observable $ \hat{X}_a $. Similarly, if we use the LO with additional phase $ \frac{\pi}{2} $, we have
    \begin{equation} \label{eq:Pm}
	\hat{P}_a^{(m)} \equiv \frac{i( \hat{a}^\dagger\hat{c} - \hat{a}\hat{c}^\dagger )}{2\sqrt{\langle \hat{c}^\dagger\hat{c} \rangle}} .
    \end{equation}
Observables $ \hat{X}_a^{(m)} $ and $ \hat{P}_a^{(m)} $ have dependence on the state of LOs. Therefore, care must be taken when we intend to investigate the statistics of the signal mode $ a $, as the output signal definitely contains the contribution from the statistics of the LO mode $ c $.

\section{\label{sec:firstcriteria}First order entanglement criteria}

Let us first briefly introduce HZ entanglement criteria \cite{Hillery2006PRL, Hillery2006PRA}, which have been known to be useful to address quantum entanglement for non-Gaussian states. If two modes $ a $ and $ b $ are separable, it was shown that the following inequalities must be satisfied 
    \begin{eqnarray}
	\textrm{first type: } & & | \langle \hat{a}^s (\hat{b}^\dagger)^t \rangle |^2 \leq \langle (\hat{a}^\dagger)^s \hat{a}^s (\hat{b}^\dagger)^t \hat{b}^t \rangle , \\
	\textrm{second type: } & & | \langle \hat{a}^s \hat{b}^t \rangle |^2 \leq \langle (\hat{a}^\dagger)^s \hat{a}^s \rangle \langle (\hat{b}^\dagger)^t \hat{b}^t \rangle ,
    \end{eqnarray}
	for any positive integers $s$ and $t$. 
If any of these inequalities are violated, the state is entangled. The simplest case may be considered  in the lowest order $ s = t = 1 $, for which the inequalities become
    \begin{eqnarray}
	\label{eq:HZ1st1} \textrm{first type: } & & | \langle \hat{a}\hat{b}^\dagger \rangle |^2 \leq \langle \hat{n}_a\hat{n}_b \rangle , \\
	\label{eq:HZ1st2} \textrm{second type: } & & | \langle \hat{a}\hat{b}\rangle |^2 \leq \langle \hat{n}_a \rangle \langle \hat{n}_b \rangle .
    \end{eqnarray}
It was suggested that the left-hand side (LHS) of the first type inequality $ | \langle \hat{a}\hat{b}^\dagger \rangle |^2 $ can be measured by boson number counting followed by 50/50 beamsplitter interaction \cite{Hillery2006PRA}, but this measurement requires a nonlocal interaction. Alternatively, it can be measured by local homodyne measurements because it can be written by expanding in terms of quadrature observables as
    \begin{equation} \label{eq:LHShomodyne}
	\langle \hat{a}\hat{b}^\dagger \rangle = \langle\hat{X}_a\hat{X}_b\rangle + \langle\hat{P}_a\hat{P}_b\rangle - i\langle\hat{X}_a\hat{P}_b\rangle +i\langle\hat{P}_a\hat{X}_b\rangle .
    \end{equation}
These four terms can be determined by measuring $ \hat{X} $ and $ \hat{P} $ at each mode $ a $ and $ b $.

However, as discussed in the previous section, when ideal homodyne measurements are not available, the measured observable $ \hat{X}^{(m)} $($ \hat{P}^{(m)} $) is different from the anticipated $ \hat{X} $($ \hat{P} $). Then, we need to reformulate entanglement criteria more rigorously incorporating the actually obtained measurement statistics. In terms of the measured observables,  Eq. (9) can be re-written as
    \begin{eqnarray} \label{eq:LHSatomic}
	& & \left| \langle\hat{X}_a^{(m)}\hat{X}_b^{(m)} + \hat{P}_a^{(m)}\hat{P}_b^{(m)} - i\hat{X}_a^{(m)}\hat{P}_b^{(m)} +i\hat{P}_a^{(m)}\hat{X}_b^{(m)}\rangle \right|^2 \nonumber \\
	& & = \frac{|\langle \hat{a}\hat{b}^\dagger\hat{c}^\dagger\hat{d} \rangle|^2}{\langle\hat{n}_c\rangle \langle\hat{n}_d\rangle} = |\langle\hat{a}\hat{b}^\dagger\rangle|^2 \frac{|\langle\hat{c}^\dagger\rangle|^2}{\langle\hat{n}_c\rangle} \frac{|\langle\hat{d}\rangle|^2}{\langle\hat{n}_d\rangle} .
    \end{eqnarray}
	
{\bf first-type criterion}---In the last identity, we factor out the terms of the LO modes $c$ and $d$ since those modes are to be prepared independently. Then by using the separability condition in the first type HZ criterion (\ref{eq:HZ1st1}) relevant to the signal fields only, we obtain
    \begin{eqnarray}
	& & \left| \langle\hat{X}_a^{(m)}\hat{X}_b^{(m)} + \hat{P}_a^{(m)}\hat{P}_b^{(m)} - i\hat{X}_a^{(m)}\hat{P}_b^{(m)} +i\hat{P}_a^{(m)}\hat{X}_b^{(m)}\rangle \right|^2 \nonumber \\
	& & \leq \langle\hat{n}_a\hat{n}_b\rangle \frac{|\langle\hat{c}^\dagger\rangle|^2}{\langle\hat{n}_c\rangle} \frac{|\langle\hat{d}\rangle|^2}{\langle\hat{n}_d\rangle} .
    \end{eqnarray}
This is a precise form of separability condition whose violation can manifest quantum entanglement of the two modes $a$ and $b$. One may thus try to detect entanglement by testing the above inequality, which then requires measurement of $ |\langle\hat{c}^\dagger\rangle| $ and $ |\langle\hat{d}\rangle| $ additionally as well as the intensities of the LOs in the right-hand side (RHS). 

To reduce experimental efforts, we may eliminate the dependence on LOs by optimizing the RHS terms in Eq. (11). That is, we find a general relation $ |\langle \hat{c}^\dagger \rangle|^2 \leq \langle \hat{n}_c \rangle $ and $ |\langle\hat{d}\rangle|^2 \leq \langle \hat{n}_d \rangle $ [See Appendix \ref{sec:bound1}]. The inequality is saturated when the sum of variances $ \langle\Delta^2\hat{X}\rangle $ and $ \langle\Delta^2\hat{P}\rangle $ satisfies the minimum uncertainty, i.e., by coherent states. Finally, we obtain an entanglement criterion, i.e., a state is entangled if the following inequality is violated:
    \begin{eqnarray} \label{eq:HZatom1st1}
	& & \left| \langle\hat{X}_a^{(m)}\hat{X}_b^{(m)} + \hat{P}_a^{(m)}\hat{P}_b^{(m)} - i\hat{X}_a^{(m)}\hat{P}_b^{(m)} +i\hat{P}_a^{(m)}\hat{X}_b^{(m)}\rangle \right|^2 \nonumber \\
	& & \leq \langle\hat{n}_a\hat{n}_b\rangle .
    \end{eqnarray}

This inequality  looks very similar to the original HZ criterion (\ref{eq:HZ1st1}). However, it must be noted that the LHS quantities are the actually measured quadrature amplitudes. Therefore, importantly, our criterion in its final form (12) does not require the knowledge on the statistics of the LO. It simply suggests to proceed with the usual way of measuring the intensity difference to obtain statistics on "quadrature amplitudes" (LHSs), which necessarily entail contributions from LOs. Our criterion indicates that one does not need the specifics of LOs. Of course, the LO state certainly affects the entanglement test because the actually measured statistics will vary according to LOs used. Below we discuss entanglement detection for various cases of LOs in practice.

\begin{figure}[t]
	\centering \includegraphics[clip=true, width=0.7\columnwidth]{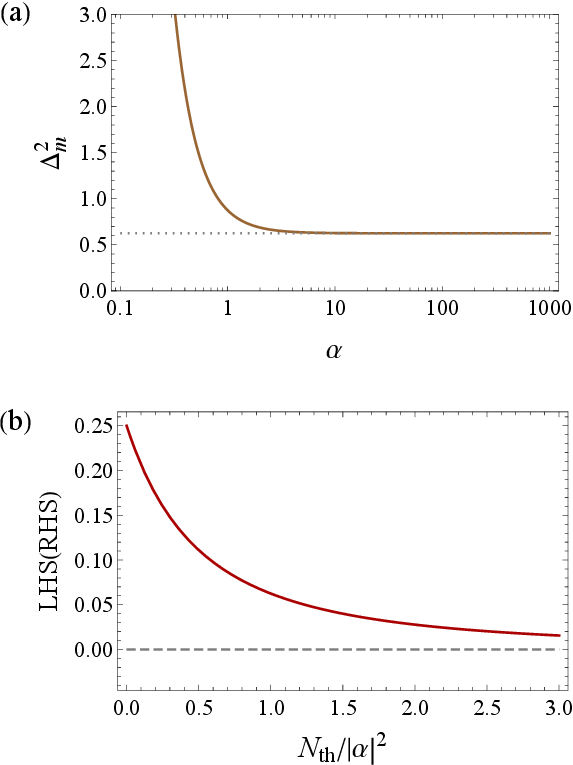}
	\caption{\label{fig:plot0110} Analysis of entanglement test for the single-excitation entangled state. (a) Plot illustrating the fluctuation $ \Delta_m^2 $ of the LHS in our criterion Eq. (12) with respect to the amplitude $\alpha$ of coherent-state LOs. Dotted line shows the fluctuation due to a finite number ($M$=1) of measurement in a large $\alpha$ limit. (b) Entanglement test using displaced thermal LOs. Red solid curve shows the LHS of Eq. (\ref{eq:HZatom1st1}) and gray dashed line shows the RHS of Eq. (\ref{eq:HZatom1st1}). }
    \end{figure}

Our criterion makes it possible to detect entangled states, which are detectable in an ideal setting by the original HZ criterion (\ref{eq:HZ1st1}), even using weak coherent-state LOs. However, if the LO amplitude is too small, it is necessary to increase the number of measurements to observe a substantial violation because the fluctuation in the measurement becomes large. As an example, we here evaluate the inequality (\ref{eq:HZatom1st1}) for the single-boson entangled state, $ \ket{\Psi_\textrm{01}} = \frac{1}{\sqrt{2}} \left( \ket{0}\ket{1} + \ket{1}\ket{0} \right) $. In this case, we see that the LHS and the RHS become $ \frac{1}{4} $ and $ 0 $, respectively, thus violating the inequality. The fluctuation of LHS is given by $ \Delta_m^2 \equiv \langle\Delta^2(\hat{X}_a^{(m)}\hat{X}_b^{(m)})\rangle + \langle\Delta^2(\hat{P}_a^{(m)}\hat{P}_b^{(m)})\rangle + \langle\Delta^2(\hat{X}_a^{(m)}\hat{P}_b^{(m)})\rangle + \langle\Delta^2(\hat{P}_a^{(m)}\hat{X}_b^{(m)})\rangle $, which we plot in Fig. \ref{fig:plot0110}(a). It is shown that $ \Delta_m $ increases as the amplitude of LOs decreases. As the precision of the measurement is given by $ \frac{\Delta_m}{\sqrt{M}} $ where $ M $ is the number of samples, $ M $ has to increase for weak LOs to make sure the violation. Note that $ \Delta_m $ converges to a nonzero value for $ \alpha \to \infty $ because the fluctuation due to a finite number of measurement is unavoidable even for the ideal homodyne measurement.

If LOs are not in coherent states, the LHS of (\ref{eq:LHSatomic}) may decrease so that the entanglement detection becomes a bit disadvantaged. The contribution from the LO identified in the LHS of Eq. (\ref{eq:LHSatomic}) is $\frac{|\langle\hat{c}^\dagger\rangle|^2}{\langle\hat{n}_c\rangle}$, which takes the maximum value of 1 for the case of coherent states. This term indicates that the LO having a coherent amplitude squared close to its intensity is desirable. However, as we have shown in Fig. 2 (a), we must take into consideration the effect of experimental fluctuation. Even for the case of coherent states that all give $\frac{|\langle\hat{c}^\dagger\rangle|^2}{\langle\hat{n}_c\rangle}=1$, the fluctuation varies with respect to the strength $\alpha$.

For instance, when LOs are in displaced thermal states $N_\textrm{th}$ with the thermal excitation, we plot the behavior of measured quadratures with respect to $N_\textrm{th}$ in Fig. \ref{fig:plot0110}(b). It is shown that the LHS decreases as $ N_\textrm{th} $ increases, but we see that the violation can always be observed owing to the RHS equal to 0. A similar behavior can be observed when we employ displaced squeezed states or displaced Fock states for LOs.

\begin{figure}[t]
	\centering \includegraphics[clip=true, width=0.7\columnwidth]{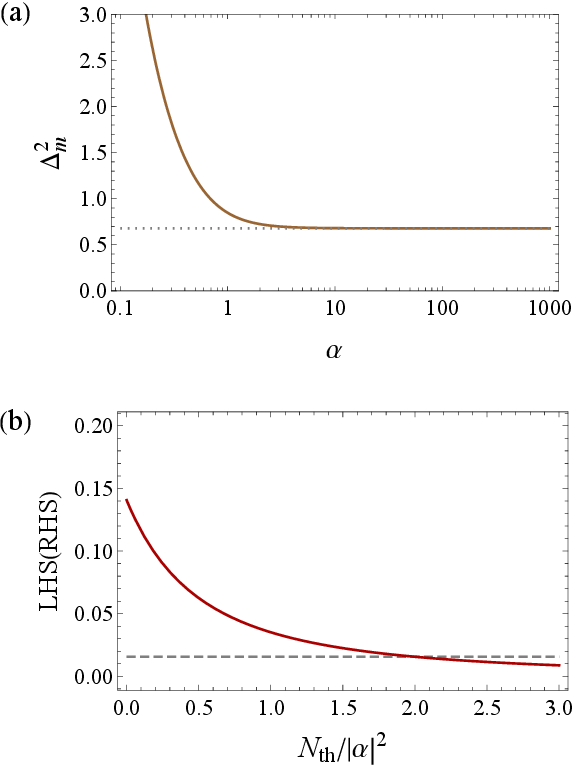}
	\caption{\label{fig:plotTMSV} Entanglement test for TMSV with 3dB of squeezing. (a) Plot illustrating the measurement fluctuation $ \Delta_m^2 $ with respect to the amplitude $\alpha$ of coherent LOs. Dotted line shows the fluctuation due to a finite number of measurement ($M$=1) in a large $\alpha$ limit. (b) Entanglement test using displaced thermal LOs. Red solid curve shows the LHS of Eq. (\ref{eq:HZatom1st2}) and gray dashed line the RHS of Eq. (\ref{eq:HZatom1st2}). }
    \end{figure}

{\bf second-type criterion}---
Now, starting from the second type HZ criterion in Eq. (8), we can obtain the entanglement criterion in a similar way, that is, a state is entangled if the following inequality is violated:
    \begin{eqnarray} \label{eq:HZatom1st2}
	& & \left| \langle\hat{X}_a^{(m)}\hat{X}_b^{(m)} - \hat{P}_a^{(m)}\hat{P}_b^{(m)} + i\hat{X}_a^{(m)}\hat{P}_b^{(m)} + i\hat{P}_a^{(m)}\hat{X}_b^{(m)}\rangle \right|^2 \nonumber \\
	& & \leq \langle\hat{n}_a\rangle \langle\hat{n}_b\rangle .
    \end{eqnarray}
This type of inequality can be violated by two-mode squeezed vacuum (TMSV) states $ \ket{\Psi_\textrm{TMSV}} = \left(1-x^2\right)^\frac{1}{2} \sum_{n=0}^\infty x^n\ket{n}\ket{n} $ where $ 0 \leq x \leq 1 $. For coherent-state LOs, the LHS and the RHS of inequality (\ref{eq:HZatom1st2}) become $ \big( \frac{x}{1-x^2} \big)^2 $ and $ \big( \frac{x^2}{1-x^2} \big)^2 $, respectively, and the violation thus occurs for any $ x > 0 $. The measurement precision $ \Delta_m^2 $ again becomes large when the size of LOs are small, as shown in Fig. \ref{fig:plotTMSV}(a). 

For displaced-thermal-state LOs, the LHS decreases as the thermal excitation number increases, which is shown in Fig. \ref{fig:plotTMSV}(b). In this case, if $ \frac{N_\textrm{th}}{|\alpha|^2} $ is too large, we fail to detect entanglement. Explicitly, the crossover occurs at $ \frac{N_\textrm{th}}{|\alpha|^2} = \frac{1-x}{x} $, which means that we need smaller $ N_\textrm{th} $ or larger $ \alpha $ to observe the violation for larger $ x $ (higher squeezing). This is because the photon number of TMSV increases as $ x $ increases so that we need LOs with a large enough amplitude.

\section{\label{sec:highercriteria}Higher order entanglement criteria}

A useful feature of HZ criteria is that they are given as a series of inequalities in different orders of field operators. For instance, we may consider the second order criterion with $ s = t = 2 $, written as
    \begin{eqnarray} \label{eq:HZ2nd}
	|\langle \hat{a}^2(\hat{b}^\dagger)^2 \rangle|^2 & = & |\langle (\hat{X}_a+i\hat{P}_a)^2 (\hat{X}_b-i\hat{P}_b)^2 \rangle|^2 \nonumber \\
	& \leq & \langle (\hat{a}^\dagger)^2\hat{a}^2(\hat{b}^\dagger)^2\hat{b}^2 \rangle .
    \end{eqnarray}
Replacing $ \hat{X} $($ \hat{P} $) with the measured observable $ \hat{X}^{(m)} $($ \hat{P}^{(m)} $), we rewrite the inequality as
    \begin{eqnarray}
	& & \left| \langle (\hat{X}_a^{(m)}+i\hat{P}_a^{(m)})^2 (\hat{X}_b^{(m)}-i\hat{P}_b^{(m)})^2 \rangle \right|^2 \nonumber \\
	& & = |\langle \hat{a}^2(\hat{b}^\dagger)^2 \rangle|^2 \frac{|\langle (\hat{c}^\dagger)^2 \rangle|^2}{\langle\hat{n}_c\rangle^2} \frac{|\langle \hat{d}^2 \rangle|^2}{\langle\hat{n}_d\rangle^2} \nonumber \\
	& & \leq \langle (\hat{a}^\dagger)^2\hat{a}^2(\hat{b}^\dagger)^2\hat{b}^2 \rangle \frac{|\langle (\hat{c}^\dagger)^2 \rangle|^2}{\langle\hat{n}_c\rangle^2} \frac{|\langle \hat{d}^2 \rangle|^2}{\langle\hat{n}_d\rangle^2} .
    \end{eqnarray}
Again if we do not have the access to the information about $ \langle (\hat{c}^\dagger)^2 \rangle $ and $ \langle \hat{d}^2 \rangle $, it is desirable to take their maximum possible values in order to eliminate the need of measuring them. Using Cauchy-Schwarz inequality $ |\langle \hat{A} \rangle|^2 = |\langle \hat{A}^\dagger \rangle|^2 \leq \langle \hat{A}^\dagger\hat{A} \rangle  $, we find
    \begin{eqnarray} \label{eq:c2bound1}
    |\langle (\hat{c}^\dagger)^2 \rangle|^2 & \leq & \langle (\hat{c}^\dagger)^2\hat{c}^2 \rangle = \langle\hat{n}_c^2\rangle - \langle\hat{n}_c\rangle , \nonumber \\ 
        |\langle \hat{d}^2 \rangle|^2 & \leq & \langle(\hat{d}^\dagger)^2\hat{d}^2 \rangle = \langle\hat{n}_d^2\rangle - \langle\hat{n}_d\rangle ,
    \end{eqnarray}
where equalities hold for coherent states. Therefore, we obtain the second order criterion given by
    \begin{eqnarray} \label{eq:HZatom2nd1}
	& & \left|\langle (\hat{X}_a^{(m)}+i\hat{P}_a^{(m)})^2 (\hat{X}_b^{(m)}-i\hat{P}_b^{(m)})^2 \rangle \right|^2 \nonumber \\
	& & \leq \langle (\hat{a}^\dagger)^2\hat{a}^2(\hat{b}^\dagger)^2\hat{b}^2 \rangle \frac{\langle\hat{n}_c^2\rangle - \langle\hat{n}_c\rangle}{\langle\hat{n}_c\rangle^2} \frac{\langle\hat{n}_d^2\rangle - \langle\hat{n}_d\rangle}{\langle\hat{n}_d\rangle^2} .
    \end{eqnarray}
In this criterion we require $ \langle\hat{n}_{c(d)}^2\rangle $ as well as $ \langle\hat{n}_{c(d)}\rangle $, which can both be obtained by measuring the particle number distribution of the LO.

An alternative bound can be derived using the positivity condition of covariance matrices [See Appendix \ref{sec:bound2}] as
    \begin{eqnarray} \label{eq:c2bound2}
        |\langle (\hat{c}^\dagger)^2 \rangle|^2 & \leq & \langle\hat{n}_c\rangle^2 + \langle\hat{n}_c\rangle , \nonumber \\
        |\langle \hat{d}^2 \rangle|^2 & \leq & \langle\hat{n}_d\rangle^2 + \langle\hat{n}_d\rangle ,
    \end{eqnarray}
where equalities hold for squeezed states. Then the second order criterion is written as
    \begin{eqnarray} \label{eq:HZatom2nd2}
	& & \left| \langle (\hat{X}_a^{(m)}+i\hat{P}_a^{(m)})^2 (\hat{X}_b^{(m)}-i\hat{P}_b^{(m)})^2 \rangle \right|^2 \nonumber \\
	& & \leq \langle (\hat{a}^\dagger)^2\hat{a}^2(\hat{b}^\dagger)^2\hat{b}^2 \rangle \frac{\langle\hat{n}_c\rangle+1}{\langle\hat{n}_c\rangle} \frac{\langle\hat{n}_d\rangle+1}{\langle\hat{n}_d\rangle} .
    \end{eqnarray}
Here we require only $ \langle\hat{n}_{c(d)}\rangle $ for LO modes. From now on, we assume that both LO modes $ c $ and $ d $ are in the same state. Whether the criterion (\ref{eq:HZatom2nd1}) or (\ref{eq:HZatom2nd2}) yields a tighter bound depends on the state of LO modes. For instance, as the inequality (\ref{eq:c2bound1}) is saturated by coherent states, the criterion (\ref{eq:HZatom2nd1}) always gives a tigher bound for coherent-state LOs. In contrast, the inequality (\ref{eq:c2bound2}) is saturated by squeezed states, and thus the criterion (\ref{eq:HZatom2nd2}) gives a tighter bound for squeezed-state LOs.

\begin{figure}[t]
      \centering \includegraphics[clip=true, width=0.7\columnwidth]{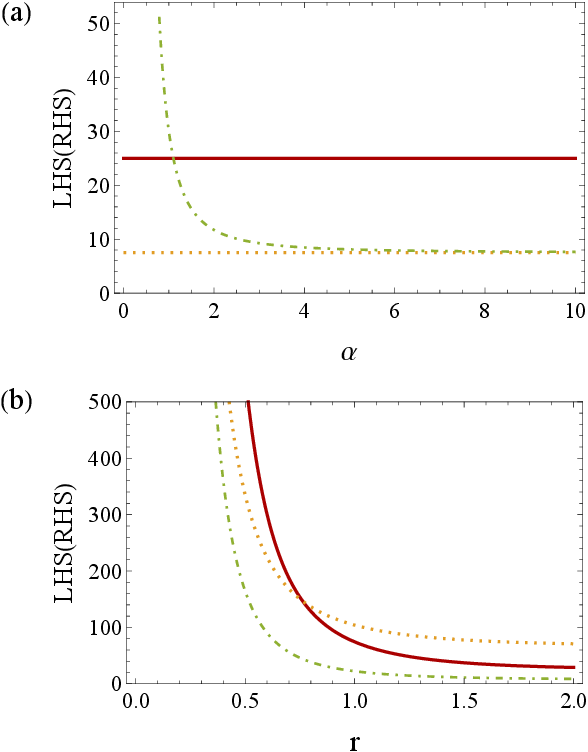}
        \caption{\label{fig:plot2nd} Entanglement test for the binomial entangled state with $ n=4 $ of total excitation using (a) coherent-state LOs and (b) squeezed-state LOs. The red solid curve shows the LHS of inequalities (\ref{eq:HZatom2nd1}) and (\ref{eq:HZatom2nd2}), and the yellow dotted curve and the green dot-dashed curve the RHSs of (\ref{eq:HZatom2nd1}) and (\ref{eq:HZatom2nd2}), respectively.}
    \end{figure}

We investigate our second order entanglement criteria for the binomial state $ \ket{\Psi_\textrm{bi}} = 2^{-\frac{n}{2}}\sum_{j=0}^n \binom{n}{j}\ket{j}\ket{n-j} $. The binomial state is widely studied in massive particle systems in the case that the total number of particles is constrained \cite{Dunningham2002,Dalton2012}. For instance, it can be generated by injecting a Fock state $|n\rangle$ and a vacuum state into a 50:50 beam splitter. In Fig. \ref{fig:plot2nd}(a,b), we compare the LHSs and the RHSs of the inequalities (\ref{eq:HZatom2nd1}) and (\ref{eq:HZatom2nd2}) employing coherent-state LOs and squeezed-state LOs, respectively.
For the case of coherent-state LO, it is shown that the criterion (\ref{eq:HZatom2nd1}) detects entanglement in all ranges of $ \alpha $, while the criterion (\ref{eq:HZatom2nd2}) cannot detect entanglement for small $ \alpha $. On the other hand, for the case of squeezed-state LO, the criterion (\ref{eq:HZatom2nd2}) detects entanglement in all ranges of $ r $, whereas the criterion (\ref{eq:HZatom2nd1}) cannot detect entanglement when the squeezing $ r $ is large. Interestingly, although the observables $ \hat{X}^{(m)} $ and $ \hat{P}^{(m)} $ obtained with squeezed-vacuum LOs show quite different statistics from $ \hat{X} $ and $ \hat{P} $, they can be successfully used to detect CV entanglement.

The condition on which criterion yields a tighter bound is fully determined by Mandel Q factor, $ Q = \frac{\langle\Delta^2\hat{n}\rangle}{\langle\hat{n}\rangle} - 1 $. Inequality (\ref{eq:c2bound1}) yields a tighter bound than inequality (\ref{eq:c2bound2}) if and only if $ Q < 1 $. Therefore, for LOs with Poissonian or sub-Poissonian distribution, one may possess an advantage in employing the criterion (\ref{eq:HZatom2nd1}) for efficient entanglement detection.

To experimentally measure the second-order quadratures properly, $ (\hat{X}^{(m)}+i\hat{P}^{(m)})^2 = (\hat{X}^{(m)})^2 + (\hat{P}^{(m)})^2 + i(\hat{X}^{(m)}\hat{P}^{(m)}+\hat{P}^{(m)}\hat{X}^{(m)}) $ in the LHS, we need to measure $ \hat{X}^{m,\phi} \equiv \frac{1}{2\sqrt{\langle \hat{c}^\dagger\hat{c} \rangle}}(e^{i\phi}\hat{a}^\dagger\hat{c} + e^{-i\phi}\hat{a}\hat{c}^\dagger) $ at three different angles $\phi$ by changing the phase of LO. One possible choice is to take $ \phi = 0, \frac{\pi}{2} $ for $ \hat{X}^{(m)} $, $ \hat{P}^{(m)} $, and $ \phi = \frac{\pi}{4} $ for $ \hat{X}^{m,\frac{\pi}{4}} = \frac{1}{\sqrt{2}}(\hat{X}^{(m)}+\hat{P}^{(m)}) $. Then $ \hat{X}^{(m)}\hat{P}^{(m)}+\hat{P}^{(m)}\hat{X}^{(m)} $ can be obtained by analyzing the statistics of those 3 observables, i.e., $ \hat{X}^{(m)}\hat{P}^{(m)}+\hat{P}^{(m)}\hat{X}^{(m)}=2 (\hat{X}^{m,\frac{\pi}{4}})^2-(\hat{X}^{(m)})^2- (\hat{P}^{(m)})^2$. Therefore, homodyne measurements with total 9 different settings (three for each mode) are sufficient to test our second order entanglement criterion.

We can also develop higher-order entanglement criteria in a similar way. In general, we need homodyne measurements with $ s+1 $ different settings to obtain the expectation values of $ s $th order quadratures by extending the above approach \cite{Wunsche1996,Park2021PRA}. Therefore, homodyne measurements with $ (s+1)(t+1) $ different settings are needed for higher order of $ s $ and $ t $.

\section{\label{sec:spinsq}Practical detection of BEC entanglement}

In this section, we demonstrate that our entanglement criterion can be applied to detect entanglement of BEC states in practical conditions. Entanglement between modes associated with different energy levels in BEC system can be generated via nonlinear interactions like the spin-changing collision \cite{Law1998,Kawaguchi2012,Stamper2013,Hamley2012,Luo2017,Kunkel2018,Lange2018}. BEC entanglement may provide useful applications in quantum technologies, especially in quantum metrology \cite{Lucke2011,Gabbrielli2015,Pezze2018}. CV entanglement between two atomic modes has been experimentally verified \cite{Gross2011,Peise2015}, where correlated atom pairs are created via collision of two atoms in a certain magnetic level acting as a pump mode. This is an analogy of optical parametric amplifier that creates two photon pairs by parametric down conversion. In optical systems, the nonlinearity is limited due to the short interaction time, whereas atomic systems do not exhibit such a restriction so that huge number of entangled pairs can potentially be created.

The spin dynamics may be restricted to the subspace with $ F = 1 $, where $ m_F = 0 $ mode plays as a pump mode and $ m_F = \pm 1 $ modes become the signal and the idler modes, respectively. The spin-changing collision in $ F = 1 $ levels is described by the Hamiltonian \cite{Law1998,Pezze2018} 
    \begin{widetext} \begin{equation} \label{eq:spin}
        \hat{H} = \lambda ( \hat{a}_1^\dagger\hat{a}_1^\dagger\hat{a}_1\hat{a}_1 + \hat{a}_{-1}^\dagger\hat{a}_{-1}^\dagger\hat{a}_{-1}\hat{a}_{-1} - 2\hat{a}_1^\dagger\hat{a}_{-1}^\dagger\hat{a}_1\hat{a}_{-1} + 2\hat{a}_1^\dagger\hat{a}_0^\dagger\hat{a}_1\hat{a}_0 + 2\hat{a}_{-1}^\dagger\hat{a}_0^\dagger\hat{a}_{-1}\hat{a}_0 + 2\hat{a}_0^\dagger\hat{a}_0^\dagger\hat{a}_1\hat{a}_{-1} + 2\hat{a}_1^\dagger\hat{a}_{-1}^\dagger\hat{a}_0\hat{a}_0 ) ,
    \end{equation} \end{widetext}
where $ \hat{a}_{m_F} $ are annihilation operators of mode $ m_F $ and $ \lambda $ is the parameter of interaction strength. In the low-depletion limit, where the pump mode is large enough compared to the number of excited atoms, the operator $ \hat{a}_0 $ can be replaced by $ \sqrt{N_0} $. In this approximation, the Hamiltonian is reduced to the typical two-mode squeezing operation and we thus obtain TMSV as an output state. On the other hand, when the population in modes $ m_F = \pm 1 $ becomes substantially large, this approximation no longer holds. A full quantum treatment of the spin-changing dynamics is studied in \cite{Law1998}, introducing angular momentum-like operators, $ \hat{L}_- \equiv \sqrt{2} ( \hat{a}_1^\dagger\hat{a}_0 + \hat{a}_0^\dagger\hat{a}_{-1} ) $, $ \hat{L}_+ \equiv \sqrt{2} ( \hat{a}_0^\dagger\hat{a}_1 + \hat{a}_{-1}^\dagger\hat{a}_0 ) $, and $ \hat{L}_z \equiv \hat{a}_{-1}^\dagger\hat{a}_{-1} - \hat{a}_1^\dagger\hat{a}_1 $. With these operators, the Hamiltonian (\ref{eq:spin}) can be written in a simple form as $ \hat{H} = \lambda( \hat{L}^2 - 2\hat{N} ) $, where $ \hat{N} \equiv \hat{a}_1^\dagger\hat{a}_1 + \hat{a}_0^\dagger\hat{a}_0 + \hat{a}_{-1}^\dagger\hat{a}_{-1} $ is the total number of atoms. For a fixed total number $ N $ of atoms , the eigenstate is given as $ \ket{N,l,m_l} $, where $ l = 0,2,4,\cdots,N $ for even $ N $, $ l = 1,3,5,\cdots,N $ for odd $ N $, and $ m_l = 0,\pm1,\pm2,\cdots,\pm l $. In the Fock-state representation, the eigenstate has the form $ \ket{N,l,m_l} = \sum_k C_k^{(N,l,m_l)} \ket{n_{-1}=k, n_0=N-2k-m_l, n_1=k+m_l} $, where the summation runs over all the states with nonnegative $ n_{-1} $, $ n_0 $, and $ n_1 $. The coefficient $ C_k^{(N,l,m_l)} $ can be determined using the property of angular momentum operators \cite{Law1998}.

\begin{figure}[t]
	\centering \includegraphics[clip=true, width=0.8\columnwidth]{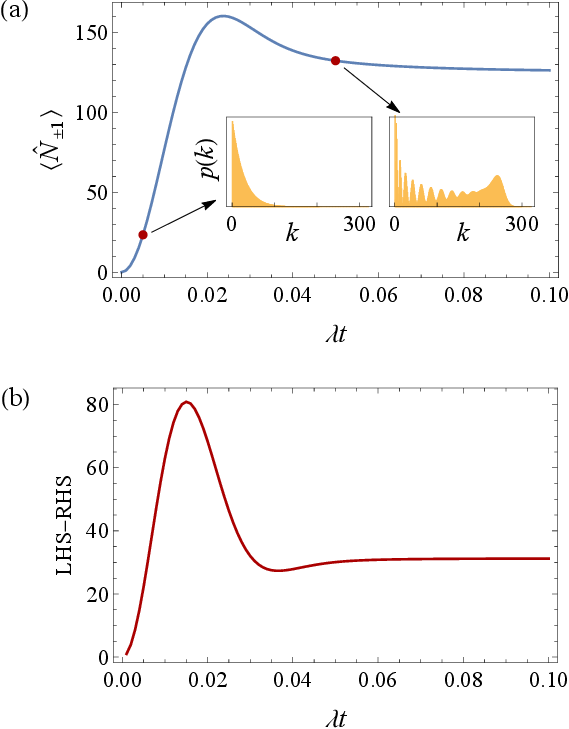}
        \caption{\label{fig:spin} (a) The population of modes $ m_F = \pm1 $ with respect to the interaction time. Insets show the atom number distribution on two specific time (marked as red dots). (b) The plot illustrating the difference between LHS and RHS of entanglement criterion (\ref{eq:HZatom1st2}).}
    \end{figure}

We start our simulation with an initial state in which $ m_F = \pm1 $ modes are empty and $ m_F = 0 $ mode is in a coherent state, i.e., $ \ket{\psi(0)} = e^{-|\alpha|^2/2} \sum_j \frac{\alpha^j}{\sqrt{j!}} \ket{0,j,0} $ in the Fock-state representation. Because the atom number difference $ n_1-n_{-1} = 0 $ is preserved, the state resides in the subspace of states $ \ket{n_{-1}=k, n_0=N-2k, n_1=k} $ with $ m_l = 0 $. In Fig. \ref{fig:spin}(a), we show the growth of the population in modes $ m_F = \pm1 $ where the initial state has the mean atom number $ \langle \hat{n}_0 \rangle = 500 $.
The population $ \langle \hat{n}_1 \rangle = \langle \hat{n}_{-1} \rangle $ increases first and then is saturated after long interaction time. The insets show the atom number distribution after short and long interaction times. In the weak interaction regime, the distribution follows that of TMSV, whereas the state becomes a nontrivial non-Gaussian entangled state after long interaction time \cite{Gross2011}. We test our entanglement criterion (\ref{eq:HZ1st2}) using coherent-state LOs and plot the difference between LHS and RHS in Fig. \ref{fig:spin}(b). It is shown that LHS is greater than RHS for $ 
\lambda t > 0 $ so that entanglement can be detected at all times even after long interaction time. This result implies that our entanglement criterion can detect spinor entangled states in a broad range of interaction strength resulting in different number/characteristics of entangled atoms.

There can be two main contributions to the statistical error. One is the intrinsic uncertainty due to a finite measurement of quantum states and the other the excess noise during the atom number counting. The error due to a finite measurement is unavoidable but the fluctuation can be reduced by increasing the number of measurement $ M $. The excess noise is mainly caused by photon shot noise in the absorption imaging \cite{Muessel2013}, which leads to an overestimation of number of atoms, say $ \Delta N_\textrm{ex} $. In the homodyne measurement, we obtain unbiased expectation values because we take the difference between atom numbers of two different modes. The excess noise only has an affect on the fluctuation such that $ \Delta X_a^{(m)} \sim \frac{\Delta N_\textrm{ex}}{\langle \hat{n}_c \rangle} $. We can thus make the fluctuation small enough with the coherent-state LO with the practically available size. For example, the achievable precision of current technology is $ \Delta N_\textrm{ex} \approx 4 $ \cite{Muessel2013,Schmeid2016} so that the coherent-state LO with $ \langle \hat{n}_c \rangle \gtrsim 100 $ is enough to make the fluctuation much smaller than the intrinsic uncertainty. On the other hand, in the atom number counting on the RHS, one needs to subtract the photon shot noise, which can be precisely determined by measuring the signal without atomic absorption \cite{Gross2011,Muessel2013,Esteve2008}.
	
\section{\label{sec:discussion}Discussion}


There have recently been growing interests in CV quantum information employing atomic systems and they require techniques for manipulating and measuring CV quantum observables appropriately. In this work we have developed entanglement criteria that are useful to address non-Gaussian entangled states for a massive system. In this case, the usual homodyne detection must be carefully analyzed incorporating the quantum statistics of the non-ideal LOs. We have particularly derived the modified Hillery-Zubairy criteria in the forms only requiring the measurement of intensity of LOs. For the case of first-order criterion, we have obtained a criterion that does not require the knowledge over the used LOs. However, the property of the LO naturally affects the measured statistics of quadrature amplitudes and we have illustrated this for the case of coherent-state, displaced-thermal state, and squeezed-state LOs, respectively. Our criteria, i.e., Eqs. (12), (13), (17) and (19), can be adopted generally for an arbitrary CV quantum system. Further studies on the CV nonclassical characteristics of massive-particle systems may pave the way towards applications of CV quantum information processing in atomic systems, such as quantum gravity sensing \cite{Howl2021} and quantum computation \cite{Fluhmann2019}. One may also pursue to develop a criterion to detect a stronger form of quantum correlation, so-called EPR correlation \cite{Peise2015,Ferris2008,Reid2009}, using atomic homodyne measurements.

In optical systems, techniques have recently been developed to enable weak-field homodyne measurements, e.g., so as to explore both wave and particle natures of quantum states \cite{Donati2014,Thekkadath2020}. It would be interesting to extend our approach in a similar context and also develop entanglement criteria that detect a various class of entangled states under a variety of LOs.

\section*{Acknowledgement}
JL and JK were supported by KIAS Individual Grants (CG073102 and CG014604) at Korea Institute for Advanced Study, respectively. J.P. acknowledges support by the National Research Foundation of Korea (NRF) Grant funded by the Korea government (MSIT) (NRF-2019R1G1A1002337 and NRF-2022M3H3A1085772). HN acknowledges the support by an NPRP grant 13S-0205-200258 from Qatar National Research Fund.

\appendix

\section{\label{sec:bound1}Bound of $ |\langle \hat{c}^\dagger \rangle|^2 $ and $ |\langle \hat{d} \rangle|^2 $}
The mean photon number of a single mode $c$ state can be written as
    \begin{eqnarray}
    \langle\hat{n}_c\rangle & = & \langle\hat{X}_c^2\rangle + \langle\hat{P}_c^2\rangle - \frac{1}{2} \nonumber \\
	& = & \langle\hat{X}_c\rangle^2 + \langle\hat{P}_c\rangle^2 + \langle\Delta^2\hat{X}_c\rangle + \langle\Delta^2\hat{P}_c\rangle - \frac{1}{2} \nonumber \\
	& = & |\langle \hat{c}^\dagger \rangle|^2 + \langle\Delta^2\hat{X}_c\rangle + \langle\Delta^2\hat{P}_c\rangle - \frac{1}{2} .
    \end{eqnarray}
The first term $ |\langle \hat{c}^\dagger \rangle|^2 $ is the contribution made out of a coherent amplitude. The remaining term $ \langle\Delta^2\hat{X}_c\rangle + \langle\Delta^2\hat{P}_c\rangle - \frac{1}{2} $ represents the contribution from variances of quadratures, which is nonnegative due to the uncertainty relation. Therefore we have $ |\langle \hat{c}^\dagger \rangle|^2 \leq \langle\hat{n}_c\rangle $ and, similarly $ |\langle \hat{d} \rangle|^2 \leq \langle\hat{n}_d\rangle $. The inequalities are saturated by states satisfying the minimum uncertainty, that is, coherent states. If LOs are not in coherent states, the term $\frac{|\langle\hat{c}^\dagger\rangle|^2}{\langle\hat{n}_c\rangle}$ becomes less than 1, which decreases the LHS of Eq. \ref{eq:HZ1st1}. For example, displaced thermal states with displacement $ \alpha $ and thermal photon number $ N_\textrm{th} $, we have $ \frac{|\langle\hat{c}^\dagger\rangle|^2}{\langle\hat{n}_c\rangle} = 1-\frac{N_\textrm{th}}{|\alpha|^2} $.

\section{\label{sec:bound2}Bound of $ |\langle (\hat{c}^\dagger)^2 \rangle|^2 $ and $ |\langle \hat{d}^2 \rangle|^2 $}
A simple calculation leads to
	\begin{eqnarray} \label{eq:cdaggersq}
	|\langle (\hat{c}^\dagger)^2 \rangle|^2 & = & | \langle\hat{X}_c^2\rangle - \langle\hat{P}_c^2\rangle + i\langle\hat{X}_c\hat{P}_c+\hat{P}_c\hat{X}_c\rangle |^2 \nonumber \\
	& = & ( \langle\hat{X}_c^2\rangle - \langle\hat{P}_c^2\rangle )^2 + \langle\hat{X}_c\hat{P}_c+\hat{P}_c\hat{X}_c\rangle^2 \nonumber \\
	& = & ( \langle\hat{X}_c^2\rangle + \langle\hat{P}_c^2\rangle )^2 + \langle\hat{X}_c\hat{P}_c+\hat{P}_c\hat{X}_c\rangle^2 \nonumber \\
	& & \quad - 4\langle\hat{X}_c^2\rangle\langle\hat{P}_c^2\rangle .
	\end{eqnarray}
Let us introduce $ a \equiv \langle\Delta^2\hat{X}_c\rangle $, $ b \equiv \langle\Delta^2\hat{P}_c\rangle $, and $ c \equiv \frac{1}{2}\langle \Delta\hat{X}_c\Delta\hat{P}_c + \Delta\hat{P}_c\Delta\hat{X}_c \rangle $, which are elements of covariance matrix $ V = \begin{pmatrix} a & c \\ c & b \end{pmatrix} $. From the uncertainty relation $ V \ge \frac{i}{4} \begin{pmatrix} 0 & 1 \\ -1 & 0 \end{pmatrix} $, the inequality $ ab-c^2-\frac{1}{16} \ge 0 $ must be satisfied. As $ |\langle (\hat{c}^\dagger)^2 \rangle|^2 $ is invariant under the phase rotation $ \hat{c} \mapsto e^{i\phi}\hat{c} $, we may assume $ \langle\hat{P_c}\rangle = 0 $. Then Eq. (\ref{eq:cdaggersq}) becomes
    \begin{eqnarray} \label{eq:c2ineq}
    |\langle (\hat{c}^\dagger)^2 \rangle|^2 & = & ( \langle\hat{X}_c^2\rangle + \langle\hat{P}_c^2\rangle )^2 + 4c^2 - 4b(a+\langle\hat{X}_c\rangle^2) \nonumber \\
    & \leq & (\langle\hat{n}_c\rangle+\frac{1}{2})^2 + 4(c^2-ab) \nonumber \\
    & \leq & \langle\hat{n}_c\rangle^2+\langle\hat{n}_c\rangle,
    \end{eqnarray}
where the inequality in the second line is saturated by states with zero first moments and the one in the last line is saturated by all pure Gaussian states. In a similar way, we obtain $ |\langle \hat{d}^2 \rangle|^2 \leq \langle\hat{n}_d\rangle^2+\langle\hat{n}_d\rangle $.

\end{document}